\documentclass[twocolumn,pre,showpacs,floatfix]{revtex4}
\pdfoutput=1
\usepackage{epsfig}
\usepackage{amsfonts}
\usepackage{amssymb}
\usepackage{amsmath}
\usepackage{pgf,pgfarrows,pgfnodes,pgfautomata,pgfheaps,pgfshade}
\usepackage{graphics,graphicx}

\begin{document}
\title{\bf Asymmetric ac fluxon depinning in a
 Josephson junction array: A highly discrete limit}
\author{Yaroslav Zolotaryuk}
\email{yzolo@bitp.kiev.ua}
\affiliation{$^1$ Bogolyubov Institute for Theoretical Physics,
National Academy of Sciences of Ukraine,
 03680 Kyiv, Ukraine}

\date{\today}

\begin{abstract}
Directed motion and depinning of topological solitons
in a strongly discrete damped and biharmonically ac-driven
array of Josephson junctions is studied. The mechanism of the depinning
transition is investigated in detail.
We show that the depinning process takes place through
chaotization of an initially standing fluxon periodic orbit.
Detailed investigation of the Floquet multipliers of these orbits
shows that depending on the depinning parameters (either the
driving amplitude
or the phase shift between harmonics) the chaotization
process can take place either along the period-doubling scenario
or due to the type-I intermittency.

\end{abstract}
\maketitle

\section{Introduction}
\label{intro}

Nonlinear dynamics of Josephson junction arrays (JJAs) has been a
subject of extensive experimental and theoretical research
\cite{u98pd,wzso96pd}.  The resistively and capacitively
shunted junction (RCSJ) model of these arrays
is described by the discrete sine-Gordon
(DSG) equation which is ubiquitous in nonlinear
physics \cite{fm96ap,bk98pr}.
Among the actively discussed problems for the JJA dynamics
the problem of the topological soliton (fluxon)
response to the time-periodic
bias, including the fluxon depinning, remains to be important.
Properties of the small rf-biased  Josephson junctions have
been extensively studied both experimentally (starting from the
pioneering papers of Shapiro \cite{s63prl}) and theoretically
(with the focus on the phase-locking \cite{k81japI} and chaotic
regimes \cite{k81japII,k96rpp}).
In particular,
the rf-biased Josephson junctions have been used as a voltage standard
\cite{volts,k96rpp}.
It is well-known \cite{u98pd,bk98pr} that a fluxon in a JJA is pinned
due to the Peierls-Nabarro (PN) potential, unless a
sufficiently strong bias is applied.
 While the depinning of nonlinear
excitations under the dc bias has been studied relatively well
(see, e.g., Refs. \cite{ucm93prb,fm96ap,fm99jpc,bhz00pre}),
the problem of the ac depinning remains scarcely studied.

The problem of ac fluxon depinning can be split into two
cases: the symmetric and asymmetric depinning. The former
case normally corresponds to the single harmonic driven systems,
particularly being  investigated in connection to the domain wall
 depinning in disordered systems
\cite{gnp03prl,npv01prl}. In the latter case, a temporarily
asymmetric (but with zero mean value) ac drive is applied.
Here asymmetric drive means that the symmetry of the driving
function is lowered, for instance, by applying a biharmonic signal.
More details will be given in the next section. This
phenomenon is based on the so-called {\it ratchet} effect, which is a
unidirectional unbiased transport induced by symmetry breaking and
nonlinearity \cite{jap97rmp,r02pr,hm09rmp}.
The mechanism of the phenomenon is based
on the breaking of the symmetries connecting orbits with opposite
velocities in the phase space \cite{fyz00prl,dfoyz02pre}
and on the phase locking of the particle dynamics to the external
drive \cite{bs00pre,bs01pre}.
The rectification due to unbiased and temporarily asymmetric drive (normally biharmonic,
consisting of a sinusoidal signal and its overtone) has been
studied theoretically \cite{w83zpb} and experimentally \cite{m90jap}
in rf-biased small Josephson junctions.
Also, the biharmonic drive has been used
for chaos supression (see, e.g., Refs. \cite{s91prb, fr94prb}).
The experimental observation of the fluxon ratchet
in Josephson junctions has been reported in several papers.
Thus, fluxon ratchet in long Josephson
junctions (LJJ) embedded in the inhomogeneous magnetic field that
emulates spatially asymmetric
sawtooth ratchet potential have been observed \cite{c01prb}.
In Ref. \cite{bgnskk05prl} the spatial asymmetry has been achieved by injecting
external bias in a way that an effective asymmetric potential is created.
The fluxon ratchet has also been observed
in a JJA where spatial symmetry was broken by introducing
spatial modulation of the coupling energy \cite{sp05prl}.
An interesting application of the fluxon ratchet effect
as a fluxon pump has been suggested in Ref. \cite{wromn99prl}.
Finally, the experimental observation of the fluxon ratchet in an annular
LJJ due to temporal asymmetry (biharmonic rf bias) has been reported
 \cite{uckzs04prl}.
It should be emphasized, that the fluxon ratchet due to temporal asymmetry
 has not yet been investigated experimentally in discrete systems.

The ratchet phenomena have already been studied theoretically
for the nonlinear excitations in different discrete media including JJAs for
both the spatial \cite{stz97pre,fmmc99epl,tmfo00pre,m07prl,cs-rs10pre}
and temporal symmetry \cite{zs06pre,mc08prl,poalk09pd}
breakings. In particular, the difference of the ratchet motion
in discrete and continuous systems has been highlighted (e.g.,
the devils staircases and the non-zero depinning threshold \cite{zs06pre}).
Nevertheless, the transition from a pinned to a running
state in ac-driven ratchets requires a special investigation.
Also, it is worthwhile to note that finding mobile fluxons in the
limit of small coupling constants ($\kappa \ll 1$) is an interesting
challenge on its own. Therefore, addressing the above questions
is the main aim of this paper.

The paper is organized as follows. In Section \ref{model}, we
present the model. In the next section it is explained how the mode-locked
solutions are computed and how their properties are investigated.
Linear stability of the single-harmonically driven JJA is briefly
discussed in Section \ref{1harm}. Asymmetric fluxon depinning under the
biharmonic drive is described in Section \ref{2harm}.
Finally,
Section \ref{conc} contains the main conclusions of the paper.

\section{The model}
\label{model}

In this paper, we discuss the dynamics of an array of parallelly shunted
and ac-biased small Josephson junctions  within the RCSJ model.
This system is described by
the ac-driven and damped discrete sine-Gordon (DSG) equation, which
can be written in a dimensionless form as follows
\begin{equation}\label{1}
{\ddot \phi}_n-\;\kappa\; \Delta \phi_n\; +\; \sin \phi_n+\alpha {\dot
\phi}_n+E(t)=0, \;  n=1,2,\ldots, N \;.
\end{equation}
Here $\phi_n$ corresponds to the phase
difference of the wave functions at the $n$th
junction, $\Delta \phi_n \equiv \phi_{n+1}-2\phi_n+\phi_{n-1}$ is
the discrete Laplacian. The coupling constant
$\kappa=\sqrt{\Phi_0/(2\pi I_c L)}$ measures the discreteness of the array,
where $\Phi_0$ is the magnetic flux quantum, $L$ is inductance of
an elementary cell, and $I_c$ is the critical current of an individual
junction. The dimensionless dissipation parameter is then
$\alpha=\Phi_0/(2\pi I_c R)$, where $R$ is the resistance of an
individual junction, and the time is normalized to the inverse
Josephson
plasma frequency $1/\omega_0=\sqrt{C\Phi_0/(2\pi I_c)}$ with $C$ being
the junction capacitance.
Finally,  $E(t)=E(t+T)$, $T=2\pi/\omega$ is an external bias, which
has a zero mean value [$\langle E(t) \rangle_t=0$],
applied to each junction of the array. In the
following, we assume $E(t)$ to be of the form
\begin{equation}\label{2}
E(t)=E_1 \cos(\omega t)+E_2 \cos (2 \omega t+ \theta),
\end{equation}
Notice that the superposition of the two harmonics makes the periodic
force to be asymmetric in
time for almost all values of $\theta$, a feature which can be
used to break the temporal symmetry of the system (more details
will be given below).
Only the circular arrays with are to be considered, therefore the
periodic boundary conditions apply: $\phi_{n+N}(t)=\phi_n(t)+2Q\pi$,
 ${\dot \phi}_{n+N}(t)={\dot \phi}_n(t)$, where
$Q$ is an integer constant that stands for the net number of fluxons trapped
in the array. Further on only the case of one fluxon ($Q=1$) will be
considered.

In JJAs the topological solitary waves have the physical meaning of
trapped magnetic flux quanta (fluxons) and the average voltage drop
reads
\begin{equation}
 V= \frac{1}{N}\sum_{n=1}^N \lim_{t
\rightarrow \infty} \frac{1}{t}\int_0^t {\dot \phi}_n(t') dt'~.
\end{equation}
If the
 fluxon is moving with a non-zero net velocity then $V \neq 0$ and
$V=0$ otherwise.

The experiments with annular JJAs have been performed for
typical lengths $N \sim 8 \div 30$ (see
Refs.~\cite{wzso96pd,u98pd,baufz00prl}). In the following, we consider
the case of an array with $N=10$ junctions subjected to periodic
boundary conditions (annular array) unless stated otherwise.

The unidirectional fluxon motion
 can take place either on  regular trajectories
(limit cycles) or on chaotic trajectories. Further on
we will refer to the trajectories with $V \neq 0$
as to the {\it transporting} ones, while the trajectories with
$V=0$ will called {\it non-transporting} trajectories.
Obviously, only the transporting trajectories are of interest within
 this paper.

The regular transporting trajectories correspond
to the limit cycles of Eq.~(\ref{1}), which are mode-locked to the
frequency of the external bias. On this orbit, the
average kink velocity is expressed as
$\langle v \rangle = \frac{k}{l}\cdot
\frac{\omega}{2\pi}$, where the winding numbers
 $k$ and $l$ are integer. In the resonant
regime, the fluxon travels $k$ sites during the time $lT=2\pi
l/\omega$, so that, except for a shift in space, its profile is
completely reproduced after this time interval (in the pendulum
analogy, this orbit corresponds to $k$ full rotations of the
pendulum during $l$ periods of the external drive).
The voltage drop in an annular JJA (with one fluxon in it) is
related to the average fluxon velocity $\langle v \rangle$ by the
equation $V=2\pi \langle v \rangle /N$ \cite{u98pd}.

According to Ref. \cite{zs06pre}, in order to obtain the
directed fluxon motion, all symmetries that relate two fluxons
with opposite velocities should be broken. It happens if the
following inequality is true: $E(t) \neq -E(t+T/2)$.
The bias (\ref{2}) satisfies it, but in the
strictly Hamiltonian case ($\alpha=0$) an additional condition
is necessary:  $E(-t+t')\neq E(t)$.

\section{Floquet analysis of the mode-locked states}
\label{floquet}
%

In order to understand better the depinning transition, we focus first on
the regular mode-locked solutions (limit cycles) of the driven DSG equation.
The fluxon periodic orbit is computed  by finding zeroes of the map
\begin{equation}\label{map}
{\hat{\cal I}_{kl}}(T) {\bf X}= {\bf X} ,
\end{equation}
where the vector ${\bf X}$ consists of the dynamical variables
$\{\phi_n,{\dot \phi}_n\}_{n=1}^N$.
The operator ${\hat {\cal I}}_{kl}$ stands for the integration of the equations
of motion (\ref{1}) during the time $l T$ and afterwards the shift of the final
solution by $k$ sites forward if $k<0$ or backward if $k>0$.
The case $k=0$ corresponds to the fluxon pinned to a lattice site.

A fixed point of the map (\ref{map}) is a mode-locked solution
$\{\phi_n^{(0)}(t),{\dot \phi}_n^{(0)}(t)\}_{n=1}^N$
which reproduces itself after the time $lT$
with the space shift by $k$ lattice sites backward or forward. Next, we
substitute the expansion
\begin{equation}
\phi_n(t)=\phi_n^{(0)}(t)+\varepsilon_n(t)~,
\end{equation}
 into Eq. (\ref{1}).
For the case of {\it standing} fluxon ($k=0$)
after keeping only the linear terms, we obtain the following set
of linear ODEs with periodic coefficients:
\begin{equation}\label{A3}
{\ddot \varepsilon}_n=-\alpha {\dot \varepsilon}_n + \kappa \Delta \varepsilon_n
-\cos [\phi_n^{(0)}(t)]\varepsilon_n~,~~ n=1,2,\ldots,N.
\end{equation}
The map
\begin{equation}
\left [ \begin{array}{c} \vec{\varepsilon}(lT) \\
{\dot {\vec \varepsilon}}(lT)\end{array} \right ]=
 {\hat M}(T) \left [ \begin{array}{c} \vec{\varepsilon}(0) \\
{\dot {\vec \varepsilon}}(0)\end{array}
\right ]
\end{equation}
 is constructed from the solutions of the system (\ref{A3}). It
relates the small perturbations
${\vec \varepsilon}(t)=\{\varepsilon_n(t)\}_{n=1}^N$ at
the time moments $t=0$ and $t=lT$.
The $2N\times 2N$ Floquet (monodromy) matrix $\hat M$ contains all the
necessary information about the linear stability of the system.
If this matrix has at least one eigenvalue with
$|\Lambda_n|>1$ ($n=1,2,\ldots, 2N$), then the system is unstable. If for
all eigenvalues
$|\Lambda_n|\le 1$, the system is stable. It is well-known \cite{via89}
that these eigenvalues
come in quadruples, so that if $\Lambda_n$
is an eigenvalue, then $\Lambda_n^*$, $R/\Lambda_n$ and $R/\Lambda_n^*$
(here $R=e^{-l\alpha \pi/\omega}$, see, for example, Refs. \cite{mffzp01pre,mfmf01pre})
are also eigenvalues.
Thus, the Floquet multipliers lie either on the circle
of the radius $R$ (will be referred to as a {\it $R$-circle}
throughout the paper) or may depart from it after collisions.
The notable difference of the ac-driven case
from the dc-driven (autonomous) case is the absence of the degeneration
with respect to time shifts, which manifests itself in the absence
of the eigenvalue $\Lambda=1$ \cite{sm97no}.

With the help of the Newton-Raphson iterative method it is possible
to compute numerically
the respective mode-locked limit cycle for the given period $lT$
 with a desired computer precision.
For details one might consult Ref. \cite{fw98pr}.
The advantage of this approach
is that not only attractors, but also repellers, can be computed.
Also, wrong conclusions which can be made due to sensitivity
to initial conditions can be avoided.

In this paper, we plan to compute the mode-locked limit
cycle that corresponds to the standing fluxon and to path-follow
it while a control parameter is changed until the cycle becomes
unstable or completely disappears. By monitoring the moduli of
the Floquet eigenvalues $|\Lambda_n|$ one can obtain the information
about the underlying bifurcations and, consequently, about the
depinning transition.

\section{Fluxon dynamics under the single-harmonic drive}
\label{1harm}
%

Before investigating the problem of the directed fluxon motion under the
influence of a biharmonic signal (\ref{2}) we briefly consider the case
of a single harmonic drive when $E_2=0$. It is of interest to investigate
the stability of a mode-locked state that corresponds to a standing fluxon.
The existence of the non-zero Peierls-Nabarro (PN) potential causes fluxon
pinning to the lattice \cite{pk84pd}.
Intuitively, it is not hard to understand that on the parameter
plane $(\kappa,E_1)$ one can draw a curve that separates
the area where only the stable mode-locked standing fluxons exist
(the pinning area) from the are where mobile fluxons, both chaotic and
regular, coexist together with the standing fluxons (the transporting area).
This curve should exist as a dependence $E_1^{(c)}=E_1^{(c)}(\kappa)$
 where for a given $\kappa$ only pinned mode-locked
states exist if $E_1<E_1^{(c)}$. If $E_1>E_1^{(c)}$, the dynamics
appears to be more complex including diffusively moving fluxons, mode-locked
moving fluxons or chaotic dynamics
of the whole array when individual fluxons cannot be identified.
The latter case corresponds to the {\it non-existing} area and will not be
discussed in this paper.
In the limit $\kappa \to \infty$, the effects of discreteness
disappear, thus $E_1^{(c)}$ should decrease. On the other hand, the decrease
of $\kappa$ means that the PN barrier becomes
stronger and thus a larger amplitude is necessary to overcome it.
As a result,  $E_1^{(c)}$  increases when $\kappa \to 0$.
The exit from the
pinning area [below the curve $E_1^{(c)}(\kappa)$] can lead
to different scenarios depending on the direction of the exit.

The issue of discrete kink unlocking (depinning) in the DSG lattice
has been studied in Ref.~\cite{mfmfs97prb}, but
 for the parameter ranges which are different from ours. We are
 dealing with large bias amplitudes and small couplings.
Especially it is useful to focus on the Floquet analysis of
pinned fluxon states in the limit of small ($\kappa \ll 1$) couplings.
Also, this section will make more clear the presentation of
the original results  in the next sections.

\begin{figure*}[htb]
\centerline{\includegraphics[width=1.98\columnwidth]{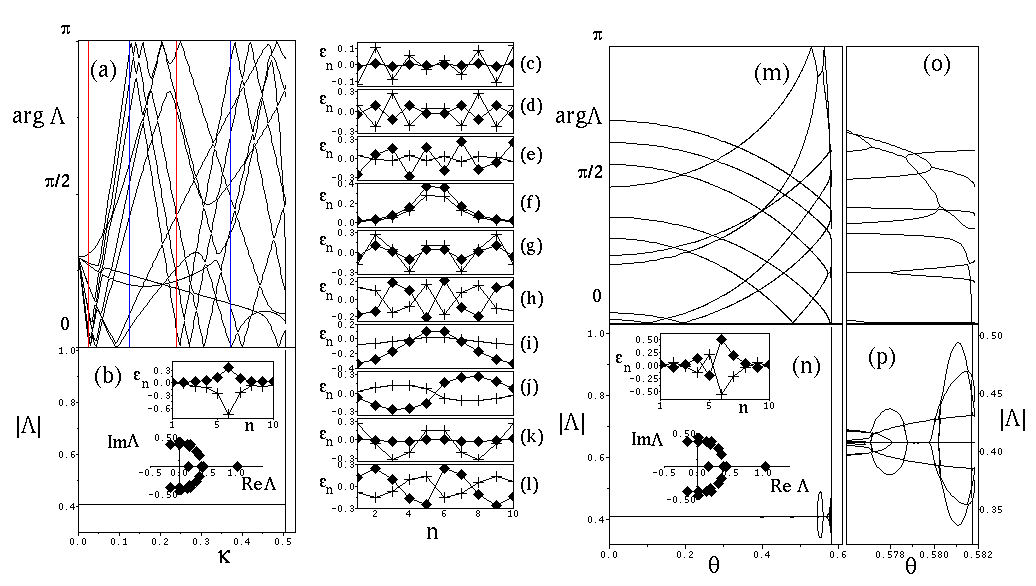}}
\caption{(Color online) Phases (a) and moduli (b) of Floquet eigenvalues
as function of $\kappa$ for $E_1=0.05$,  $\alpha=0.1$, and
$\omega=0.35$.
Floquet eigenvectors [panels (c)-(l)] at $E_1=0.05$, $\kappa=0.1$
(see text for details).
Panels (m)-(p) show the phases and moduli
of  Floquet eigenvalues at $\kappa=0.1$ as function of $E_1$.
Upper inset in panel (b) shows the eigenvector for  unstable eigenvalue
at $\kappa=0.5058245$
[${\mbox Re}~\varepsilon_n$ ($+$) and
${\mbox Im}~\varepsilon_n$ ($\blacklozenge$)]. Same for the inset in panel
(n) at $E_1=0.58183325$.
The lower insets shows the positions of eigenvalues.}
\label{fig1}
\end{figure*}

The behavior of the Floquet eigenvalues (computed as described in
Sec. \ref{floquet}) is shown in detail in Fig.~\ref{fig1}. In
panels (a)-(b), their evolution as a function of
the coupling constant $\kappa$ is given starting from the
anticontinuum ($\kappa=0$) limit.
%
Note that the eigenvalues are placed on the complex plane
symmetrically with respect to the real axis, thus we limit ourselves
with the eigenvalues lying in the upper half-plane ($\mbox{Re}\Lambda \ge 0$).
The bias amplitude now is $E_1=0.05$, so that  the excitations
on the fluxon background can be treated as small.
In the anticontinuum limit, all the eigenvalues sit in one point
and with the growth of $\kappa$ they separate forming two distinct
groups: the modes associated with the linear spectrum (Josephson plasmons)
and the internal mode(s). The plasmon band extends linearly with $\kappa$
according to the dispersion law
\begin{equation}\label{wl}
\omega_L(q)=\sqrt{1+4\kappa~ \sin^2\frac{q}{2} }~.
\end{equation}
Due to finiteness of the array, the wavenumber $q \in [0,2\pi)$ attains
only discrete set of values $q_m=2\pi m/N$, $m=\pm 1,\ldots  , \pm N$.
The localized eigenmode detaches itself from the linear band: as
it is well known in the discrete Klein-Gordon theory \cite{bkm98pd},
the internal mode is soft, so that it detaches from the linear band into the gap.
In  panels (c)-(l) of Fig.~\ref{fig1}, the shapes of the eigenvectors
  ${\vec \varepsilon}$
(both real and imaginary parts) are shown for the parameter values
$E_1=0.05$ and $\kappa=0.1$. These eigenvectors are ordered from top to
bottom according to the decrease of $\arg \Lambda_n$.
From the shape
of the eigenvectors one can conclude that all of them, except one,
are delocalized
and therefore they are associated with the linear spectrum.
The eigenmode in Fig.~\ref{fig1}(f) has a clear localized structure.
In our case, the fluxon state is locked to the external drive with
the frequency $\omega=0.35$ which lies in the gap of the spectrum
 (\ref{wl}). Only the overtones $n\ge 3$ can lie in the linear
spectrum: $1<n\omega<\sqrt{1+4\kappa^2}$, therefore the excited linear modes
satisfy the condition $n\omega=\omega_L(q_m)$, $m=1,2,\ldots, N$. The
increase of $\kappa$ widens the
linear spectrum and, as a result, the respective eigenvalues spread
around the $R$-circle.
The collisions of the eigenvalues
at $\arg \Lambda=\pi$ with the highest ($m=N$) linear mode
are marked by the vertical red lines at $\kappa=[(7\omega/2)^2-1]/4$ and
$\kappa=[(9\omega/2)^2-1]/4$. They correspond to the parametric resonances
of this mode with the external drive: $\omega_L(q_N)=7\omega/2$ and
$\omega_L(q_N)=9\omega/2$. Collisions at
 $\arg \Lambda=0$
 marked by the blue lines at $\kappa=[(3\omega)^2-1]/4$
and $\kappa=[(4\omega)^2-1]/4$ and correspond to the main resonances
$\omega_L(q_N)=4\omega$ and $\omega_L(q_N)=4\omega$.
The internal mode evolution with varying $\kappa$ can be clearly seen due
to its nonlinear behavior as a function of $\kappa$, as compared
to the linear evolution of plasmon modes.
After interactions with the linear
spectrum (marked by the avoided crossings), the internal mode hits the
real axis at  $\arg \Lambda=0$,
thus signaling the resonance with some overtone of the driving
frequency. After that the eigenvalue increases extremely fast
[note the almost vertical growth of $|\Lambda_n|$ in Fig.~\ref{fig1}(b)],
leaves the $R$-circle and finally exits the unit circle. We conclude that
the disappearance of the standing mode-locked fluxon
takes place via the tangential (saddle-node) bifurcation. The excitation of the
fluxon along the destabilizing direction in the phase space leads to
{\it chaotic} fluxon diffusion with the net zero velocity.

While the previous investigation took place at rather small amplitudes
for which the excitation on the fluxon background can be considered as
small, now we focus on the
evolution of the Floquet eigenvalues as a function of the driving
amplitude for the fixed value of coupling $\kappa=0.1$.
As one can see in Fig. \ref{fig1}(m-n),
the eigenvalue that corresponds to the localized mode moves on the $R$-circle
in the direction opposite to the motion of the linear band.
 After following
the respective eigenvalue as a function of $E_1$, we see that
 it collides with its complex conjugated counterpart
at $\arg \Lambda=\pi$ (see the resulting ``bubble'' in the $|\Lambda(E_1)|$
dependence), continues
its motion along the $R$-circle, then interacts with the eigenmodes
of the linear
spectrum and finally collides with its complex conjugate counterpart
on the real axis but at $\arg \Lambda=0$. Next, the modulus
of this eigenvalue grows fast and finally exceeds the unit circle.
Again, the standing mode-locked fluxon state disappears via the
tangential bifurcation.
The unstable eigenvector is again spatially localized (see the inset in
Fig. \ref{fig1}n).
The perturbation of the kink in
the unstable direction leads to the growth of a localized
 oscillation on the kink background that results in the kink-antikink
pair birth and eventual destruction of the one-soliton state.
Other scenarios are also possible.
Similar investigation for the case $\kappa=0.2$ shows that the system can
undergo additional bifurcations and a new standing mode locked
fluxon can appear with a slightly different shape. However, the final
disappearance of the standing kink goes similarly to the
above scenario via the tangential bifurcation. In this case, the time
evolution of the unstable fluxon turns into chaotic diffusive
jumps of the kink backward and forward with the net zero velocity.

The main conclusion of this subsection is that standing
mode-locked fluxon states disappear via a tangential bifurcation,
when the driving amplitude $E_1$ or the coupling constant
exceeds some critical value. The resulting dynamics can either
lead to fluxon chaotic diffusion or to chaotic motion of the whole
lattice and consequent fluxon destruction.
Our analysis with $N=20$ and $N=30$
has shown that the size effects are insignificant: the critical
values at which the tangential bifurcation takes place differ very
slightly. The instability evolves
via the excitation of a localized internal mode. This is in
accordance with the previous results \cite{sq02pre,m-mqms03prl,zs06pre} on the
internal mode role in the kink mobility.
More details on the kink response to the single harmonic
ac drive (including the kink dynamics)
can be found in Ref.~\cite{mfmfs97prb}. In particular, it has been
shown that in the unlocking limit the kink dynamics shows the type-I
intermittency.

\section{Asymmetric fluxon depinning }
\label{2harm}
%

Now we focus on the main task of this paper, namely on the
studies of the fluxon motion when the bias (\ref{2}) is biharmonic ($E_2\neq 0$).
Here and further on, we take $E_1=E_2$ unless stated otherwise.
According to the previous studies \cite{zs06pre,cs-rs10pre}, the fluxon
motion becomes unidirectional due to the symmetry breaking and its average velocity
depends solely on the system parameters. Also, from those papers we
already know that critical parameter values should exist
above which the applied external bias overcomes the forces of pinning
and the fluxon starts to propagate. In particular, such critical
values exist  for the amplitudes of the external
drive $E_{1,2}$ and the phase shift $\theta$.

\begin{figure}[htb]
\includegraphics[width=1.0\columnwidth]{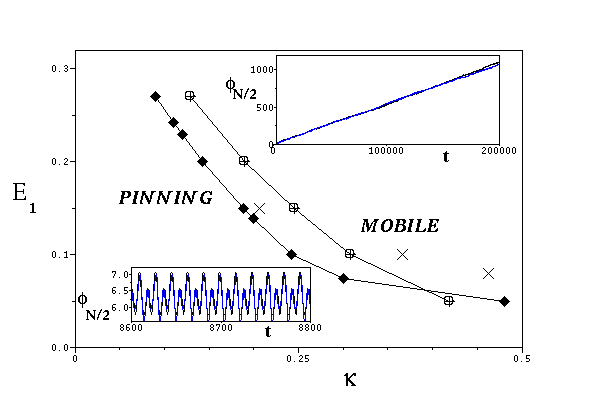}
\caption{(Color online). Existence diagram of moving fluxons on the
plane $(\kappa, E_1)$ at $\alpha=0.1$ for $\omega=0.25$ ($\oplus$),
$\omega=0.35$ ($\blacklozenge$) and $\omega=0.5$ ($\times$).
Solid lines are used as a guide for an eye
(see text for details).
The upper and lower insets show the time evolution of the central
junction for $\theta=2.5$, $\omega=0.35$, $\alpha=0.1$ and
$\kappa=0.12$, $E_1=E_2=0.22$ (lower inset) and
$\kappa=0.09$, $E_1=E_2=0.27$ (upper inset).
Different colors correspond to  different initial times:
$t_0=0$ (black) and $t_0=\pi/\omega$ (blue). }
\label{fig2}
\end{figure}

Including the second harmonic in the bias (\ref{2}), we add the
third parameter (the phase shift $\theta$)
into consideration and consequently we must think about the
existence diagram in the three-dimensional parameter space
$(\kappa,E_1,\theta)$. However, since the average voltage drop satisfies
$V(\theta)=V(\theta+2\pi)$, it is possible to construct the existence
diagram on the plane $(\kappa, E_1)$. This diagram
for the existence of moving fluxon states is shown
in Fig.~\ref{fig2}.
The numerical experiment has been performed as follows:
the pinned state with the fluxon oscillating around its center of
mass has been continued with small increase of $E_1$ or $\kappa$
until this state loses stability and starts to propagate.
 Transporting trajectories
are mostly chaotic, but regular mode-locked trajectories may also exist.
These trajectories will be analyzed in detail in the next subsections.
As one can see, similarly to the case of the single harmonic drive,
there exists a curve $E^{(c)}_1=E^{(c)}_1(\kappa)$ (
shown by markers) that separates the pinned and transporting parts
of the diagram. In the area down and to the left from the
curve $E^{(c)}(\kappa)$, there
are only mode-locked standing fluxons and no
mobile fluxons exist ($V \equiv 0$) for any $\theta \in [0,2\pi)$.
The area to the right side of the curve corresponds to the situation
where moving fluxons
%
($V\neq 0$) exist for some values of $\theta$. Obviously, in the continuum
limit $\kappa \to \infty$, the
critical depinning amplitude tends to zero, while in the anticontinuum
limit $\kappa \to 0$, it is necessary
to increase the driving amplitude in order to unlock the fluxon.
At some point the  critical drive is too strong and the
fluxon is destructed due to the fluxon-antifluxon pair creation and
the chaotic dynamics of the whole lattice. An example of
such a pinned mode-locked trajectory is shown in the lower
inset of  Fig.~\ref{fig2}.
Note that the pinning to the lattice is independent on the initial conditions.
For example, in this figure two trajectories with  the initial times $t_0=0$
and $t_0=T/2$ are shown.
Also, the scenario is the same for other values of initial time
$t_0 \in [0,T)$, the initial fluxon position, and different initial kicks.
The examples of transporting chaotic trajectories  with different initial times
($t_0=0$ and $t_0=T/2$) are illustrated by the upper inset of
 Fig.~\ref{fig2}.
Independently of the initial conditions at $t \to \infty$,
the system settles on the chaotic
attractor that corresponds to the directed fluxon motion with the
same average velocity (in this case $V \simeq 0.0055$).

Note that the principal characteristics of the $E^{(c)}_1(\kappa)$ dependence
remain qualitatively the same for different values of $\omega$.
It has been shown in Ref.~\cite{zs06pre} that the ratchet fluxon motion
is well pronounced in the frequency
range $0 \le \omega \lesssim \omega_L(0)/2$.
The Peierls-Nabarro frequency (which equals the frequency
of the internal mode
\cite{bkp97pre}) at $\kappa=0.1$ (according to the relation
$\omega_{PN}\simeq \sqrt{2\pi \alpha_0 \kappa^{3/2}}
\exp {(-\pi^2\sqrt{\kappa}/2)}$, $\;\; \alpha_0 \simeq 30 \pi \;$, see
Ref.~\cite{im82jpsj})
equals $\omega_{PN}\simeq 0.91$. Thus, for small couplings
the working frequency range lies below all characteristic system frequencies.
Among the other system parameters the coupling constant and dissipation depend
on the physical characteristics of the specific array and therefore they
cannot be changed easily in the experiment. Consequently, we are left with
the bias parameters $E_1$ and $\theta$ which can be tuned at will. Therefore it
would be
of interest to study the depinning process as a function of $\theta$ (with other
parameters fixed) and the same for $E_1$.
The following subsections will be devoted to this task.

\subsection{Depinning as a function of the bias amplitude}
\label{E1}

For continuous ratchets it was shown previously \cite{sz02pre} that the
average kink velocity is proportional to $E_1^2E_2$, so that,
provided the respective symmetries are broken,  the directed
motion can occur for arbitrary small values of the driving
amplitudes. For the DSG equation, the dependence of the average
kink velocity on the driver amplitudes is not a
smooth monotonic function of $E_{1,2}$, but a complex
function that either entirely or partially consists of
 plateaux of different lengths that correspond to mode-locked fluxon
states. At $\kappa \gtrsim 1$ this dependence resembles closely a ``devil's
staircase'', while this resemblance disappears when $\kappa$
is decreased leaving only isolated islands of regular motion \cite{zs06pre}.

\begin{figure*}[htb]
\centerline{\includegraphics[width=1.9\columnwidth]{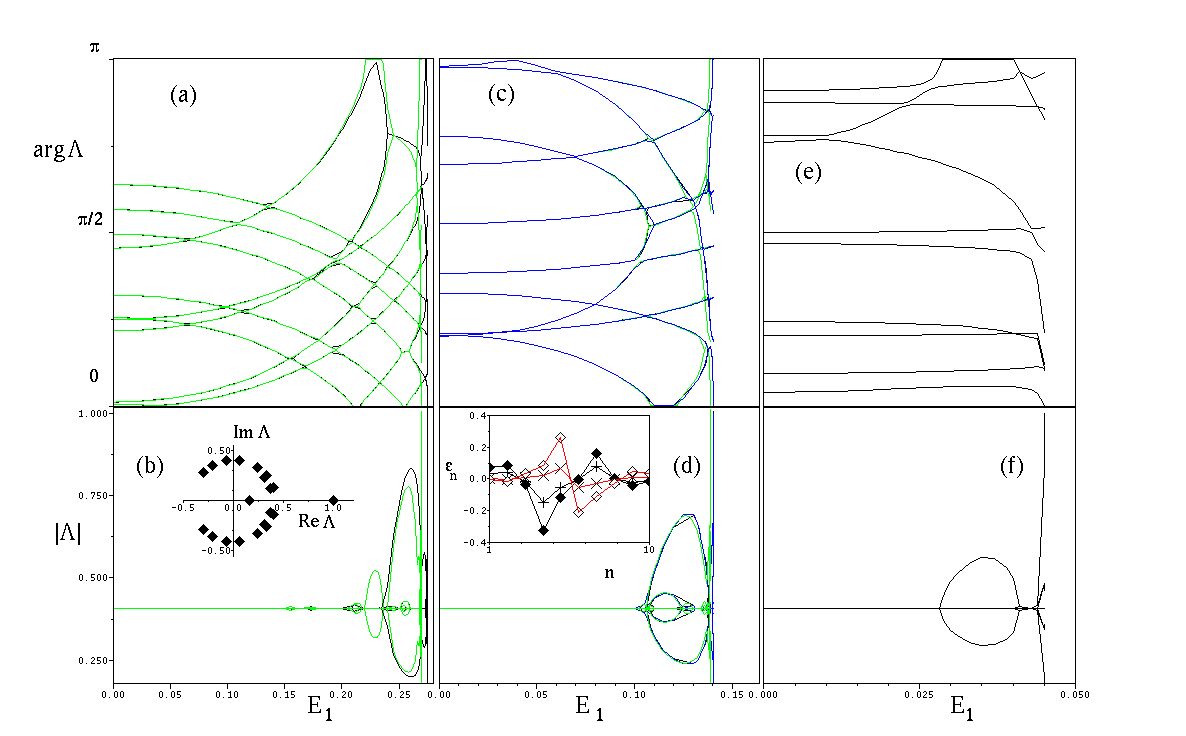}}
\caption{(Color online).
Dependence of the phases (upper figures) and moduli (lower figures) of
Floquet eigenvalues on bias amplitude $E_1$ at
$\alpha=0.1$, $\omega=0.35$, and for different coupling constants and $\theta$:
$\kappa=0.09$ (a-b), $\theta=1.5$ (black lines), $\theta=0$ (green lines);
$\kappa=0.2$, (c-d), $\theta=1.5$ (black lines), $\theta=0$ (green lines);
$\theta=-1.64159$ (blue lines);
$\kappa=0.5$, $\theta=1.5$ (e-f).
The inset in panel (b) shows the position of Floquet eigenvalues
on complex plane at $E_1=0.04511641$.
The inset in panel (d) shows the profile of  unstable
eigenvectors $\mbox {Re} \varepsilon_n$ ($+$)
and $\mbox {Im} \varepsilon_n$ ($\blacklozenge$) at
$E_1=0.1403205$, $\theta=1.5$
for eigenvalue $\Lambda=1.005497$ (black lines) and $E_1=0.14032053$,
$\theta=-1.64159$ (scaled by $2$, red lines) for  eigenvalue
 $\Lambda=1.000515$ [$\mbox {Re} \varepsilon_n$ ($\times$)
and $\mbox {Im} \varepsilon_n$ ($\diamond$)]. Solid lines are used as guides for an eye.
}
\label{fig3}
\end{figure*}

In Fig.~\ref{fig3}, the dependence of the Floquet eigenvalues  on the
driving amplitude is shown for different values of $\kappa$.
%
Comparing Figs.~\ref{fig1}(m-n) and \ref{fig3}(a-b), we observe the
close similarity in the Floquet eigenvalue behavior in the
single harmonic and biharmonic biased cases.
It appears also that independently on the coupling strength, the loss
of stability of the limit cycle appears through a collision of two
eigenvalues at $\arg \Lambda=0$ and the departure of one of the
 eigenvalues out of the unit circle [see the inset in Fig.~\ref{fig3}(b)]
that signals a tangential bifurcation taking place at some critical
value of the driving amplitude. This critical value decreases with the
growth of $\kappa$. Note that the critical  value of the bias amplitude
is approximately two times smaller than in the single harmonic case because
now the bias consists of two harmonics with the amplitude $E_1$.
In panels (a-b), it should be noted that before the tangential
bifurcation several period-doubling bifurcations take place that
can easily be identified by the bubble-like deviations on the
dependence $|\Lambda_n(E_1)|$ from the value
$|\Lambda_n|=R$. The amplitude of these deviations does
not exceed the value $|\Lambda|=1$ and they decrease with $\kappa$,
so that the limit cycle remains stable.
At higher $\kappa=0.2$, the largest bubble corresponds to a pair of
eigenvalues colliding away from the real axis (that corresponds to
the Hopf bifurcation). For even higher $\kappa=0.5$, again a bubble associated
with the period-doubling bifurcation can be seen. Although for the given range of
parameters
these bifurcations do not cause instability of the limit cycle, the increase
of $\omega$ or decrease of $\alpha$ may change the nature of the depinning
 transition. The
 computations have been performed for different values of $\theta$ and
the dependencies $\Lambda_n(E_1)$ appear to be rather similar since the curves
of different colors almost coincide. The only difference lies in the different
critical values of $E_1$, when the tangential bifurcation takes place, however
these differences are rather small.

The unstable eigenvector, shown in the inset of panel (d)
has a pronounced localized asymmetric shape. The eigenvectors that
correspond to the opposite (with respect to the shift $\theta \to
\theta \pm \pi$) values $\theta=1.5$ and  $\theta=1.5-\pi \simeq -1.64159$
are  related to each other approximately by the inversion
${\bf \varepsilon} \to -C{\bf \varepsilon}$, where $C$ is an arbitrary
constant that appears because Eqs.~(\ref{A3}) are linear. This means that
the unstable perturbation drives the fluxon
in the opposite directions for $\theta$ and $\theta+\pi$, respectively.

\begin{figure}[htb]
\begin{center}
\includegraphics[width=0.99\columnwidth]{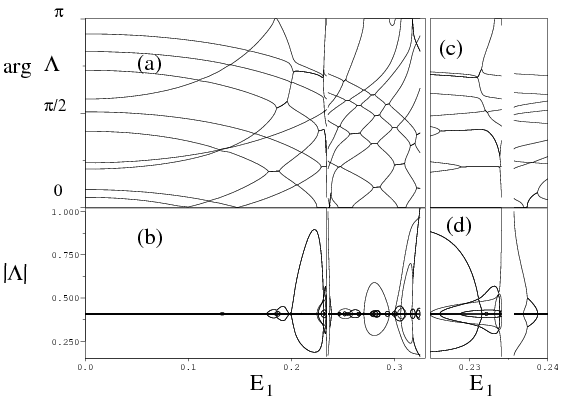}
\end{center}
\caption{
Phases (a) and moduli (b) of Floquet eigenvalues as
a function
for $\omega=0.35$, $\alpha=0.1$, $\kappa=0.12$, $\theta=1.5$. Panels (c) and
(d) give details in interval $E_1 \in $ $(0.22,0.24)$.}
\label{fig4}
\end{figure}

The behavior of the eigenvalues in the parameter space $(\kappa,E_1,\theta)$
has very complicated structure. In particular, it is possible that
for some values of $\theta$, the standing fluxon states can exist
after the tangential bifurcation shown in Fig.~\ref{fig3}. For the
sake of brevity, we are going to refer to it as to the {\it first}
tangential bifurcation. Indeed,
such an example is given by Fig.~\ref{fig4}. After the firts tangential
bifurcation, the standing solution (limit cycle) becomes
unstable at $E_1 \simeq 0.2357$ and soon disappears but reappears
back after another tangential bifurcation. In the short interval
between these bifurcations, a chaotic moving fluxon appears.
The further increase of the driving amplitude leads to the Hopf bifurcation
(collision of two eigenvalues on the complex plane) and to the eventual
departure of two eigenvalues out of the unit circle. This
instability leads to formation of a large localized excitation on the fluxon
and the eventual creation of  fluxon-antifluxon pairs. Thus, it is
possible to pass from the pinned area to the non-existing
area (where no fluxons exist due to chaos)
omitting the transporting area (where the moving fluxons exist).

%

This bifurcation analysis explains the non-monotonous behavior of the
average velocity dependence on the bias amplitude for small $\kappa$
(see Fig.~8 of  Ref.~\cite{zs06pre}), when the intervals of
zero and non-zero fluxons velocities interchange with the growth of $E_1$.


\subsection{Depinning as a function of the phase shift}

\begin{figure}[htb]
\includegraphics[width=1.0\columnwidth]{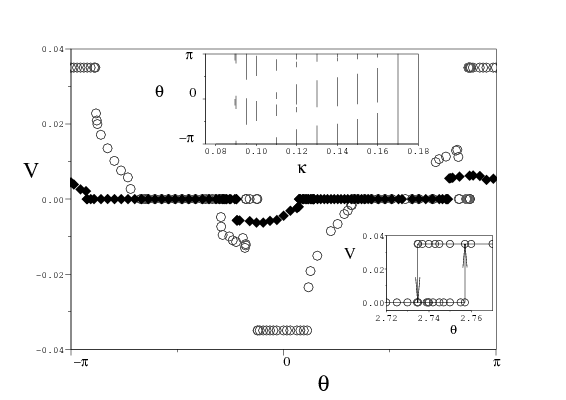}
\caption{
Average voltage drop as a function of phase shift $\theta$
for $\omega=0.35$, $\alpha=0.1$, $E_1=E_2=0.27$,
$\kappa=0.09$ ($\blacklozenge$) and $\kappa=0.12$ ($\circ$).
The upper inset shows
the existence diagram of transporting trajectories (vertical bars)
on plane $(\kappa,\theta)$. The lower inset shows the details of
hysteresis (see text).}
\label{Vth}
\end{figure}

The directed fluxon transport  in the
continuous sine-Gordon model shows \cite{sz02pre} that the dependence
of the average fluxon velocity (and, consequently, the voltage drop)
behaves as $\sin [\theta-\theta_0(\omega,\alpha)]$.
As it has been shown in Ref.~\cite{zs06pre},
in the weakly
discrete case, the two values of $\theta$ where $V(\theta)=0$ become intervals,
and the size of these intervals increases when $\kappa$ decreases.
Moreover, the dependence $V(\theta)$ changes from the
continuous behavior into the complex piecewise
function which may include resonant plateaux $V=\omega k/(lN)$ and
loses its resemblance with the sine function as long as $\kappa \to 0$.
For $\kappa \gtrsim 1$ there exist at least two intervals
on the $\theta$ axis which we denote as $(\theta^-_1,\theta^-_2)$
and $(\theta^+_1,\theta^+_2)$ where $V \neq 0$. In particular,
$V(\theta)<0$ if $\theta \in (\theta^-_1,\theta^-_2)$
and $V(\theta)>0$ if $\theta \in (\theta^+_1,\theta^+_2)$.
It appears, however, that for small $\kappa$ the dependence $V(\theta)$ may
take a more complicated shape (for example, see Fig.~8 of Ref.~\cite{zs06pre}).
This also can clearly be seen in the main graph
of Fig.~\ref{Vth}, where
the dependencies for $V(\theta)$ for several values of $\kappa$ and
for the fixed value of $E_1=0.27$ are given.
This figure in some sense is complementary to Fig.~\ref{fig2} because
it explains in detail what is happening around the curve $E^{(c)}_1(\kappa)$
when $\theta$ changes.
While at $\kappa=0.09$ only two transporting intervals where $V \neq 0$ are observed
with $\theta^-_1 \simeq -0.69$,  $\theta^-_2 \simeq 0.23$ and
$\theta^+_1 \simeq 2.47$,  $\theta^+_2 \simeq -2.92$. For other
values, namely, for $\kappa=0.12$ there can exist two
additional transporting intervals.
The transporting intervals (shown by vertical bars
on the upper inset of Fig.~\ref{Vth}) decrease as $\kappa \to 0$
at $E_1=\mbox{const}$. The same behavior occurs if $\kappa$ is fixed
and $E_1$ decreased.
%
It is worth to notice the existence of regular
transporting limit cycles that correspond
to the constant voltage plateaux $V=\pm 0.035$ for $\kappa=0.12$.
Their existence has been obtained by the direct Runge-Kutta integration
of the equations of motion and also has been confirmed by the Newton
method  with $k=\pm 1$, $l=1$. The careful
investigation of the transition between the pinned state and these
plateaux shows the existence of  narrow hystereses windows as
shown in the lower inset of Fig.~\ref{Vth}.
In order to understand the nature of the additional pinning intervals,
 the Floquet analysis of the standing
limit cycles has been performed as a function of $\theta$.

\begin{figure}[htb]
\includegraphics[width=1.0\columnwidth]{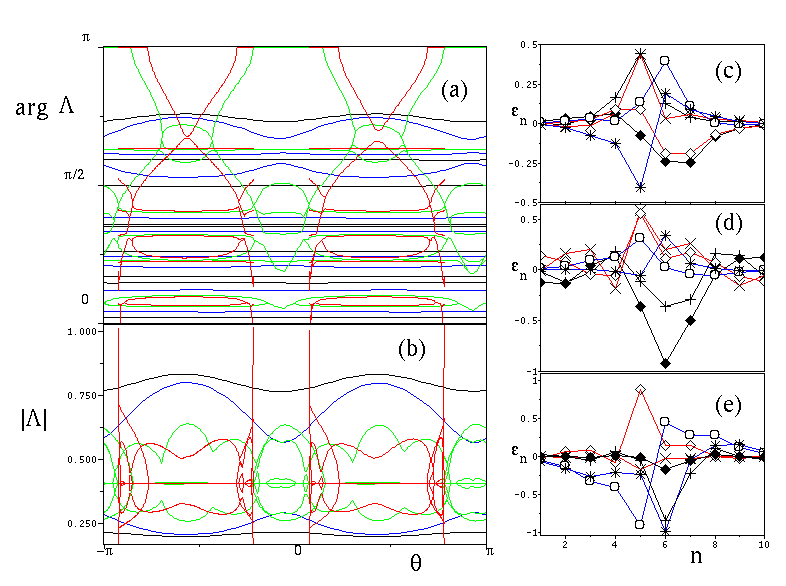}
\caption{(Color online). Phases (a) and moduli (b) of  eigenvalues
of Floquet  matrix
as a function of  phase delay $\theta$ for $\omega=0.35$, $\kappa=0.09$
and $E_1=E_2=0.26$ (black), $E_1=E_2=0.265$ (blue), $E_1=E_2=0.269$ (green)
and $E_1=0.27$ (red).
Panels (c)-(e) show the real and imaginary
parts of destabilizing eigenvectors  the eigenvalues
that lie out of the $R$-circle (see text for details):
(c) $E_1=0.26$ with $\theta=-\pi$ [red, $\mbox{Re}~\varepsilon$ ($\times$),
$\mbox{Im}~\varepsilon$ ($\diamond$)], $\theta=-\pi/2$ [blue,
$\mbox{Re}~\varepsilon$ ($*$),
$\mbox{Im}~\varepsilon$ ($\circ$)]
and $\theta=0$ [black,$\mbox{Re}~\varepsilon$ ($+$),
$\mbox{Im}~\varepsilon$ ($\blacklozenge$)];
(d) $E_1=0.269$ (the same);
(e) $E_1=0.27$ with $\theta=-2.914$ [red, $\mbox{Re}~\epsilon$ ($\times$),
$\mbox{Im}~\varepsilon$ ($\diamond$)], $\theta=-\pi/2$ [blue,
$\mbox{Re}~\varepsilon$ ($*$),
$\mbox{Im}~\varepsilon$ ($\circ$)] and $\theta=-0.69206$ [black,
$\mbox{Re}~\varepsilon$ ($+$),
$\mbox{Im}~\varepsilon$ ($\blacklozenge$)].
}
\label{eval1}
\end{figure}

In Fig.~\ref{eval1} the behavior of the Floquet eigenvalues as a
function of the
desymmetrizing parameter $\theta$ for different values of $E_1$ below
and above the depinning threshold is shown.
The left panels (a-b) correspond to the situation when
$\kappa=0.09$ and $E_1$ is increased from  below the
critical depinning dependence $E^{(c)}(\kappa)$. One can see that
initially the curves $\arg \Lambda_n(\theta)$ do not cross
each other. For $E_1=0.26$ all the eigenvalues
lie on the $R$-circle except for the two pairs with $\arg \Lambda_n$ oscillating
around $\pm 3\pi/2$ and having $|\Lambda_n| \neq R$, as it
can clearly be seen from Fig.~\ref{eval1}(a-b). These eigenvalues correspond
to the spatially localized eigenvectors as shown in  Fig.~\ref{eval1}(c).
Moreover, the non-localized eigenvalues are virtually independent on
$\theta$. The shape of the localized eigenvectors
 does not differ significantly for $\theta=\pm \pi$ and $\theta=0$ as
 well. A
slight increase of the driving amplitude to the value $E_1=0.265$ leads to
further increase of the oscillation of localized eigenvalues.
%
%
Upon the further increase of $E_1$, the eigenvalue collisions
at $\Lambda=-1$ and out of the real axis take place (see, for
example, the green line in Fig.~\ref{eval1}(a) that corresponds to
$E_1=0.269$). The monitoring of the eigenvectors that correspond to the
eigenvalues with $|\Lambda_n|\neq R$ (see panel (d) of Fig.~\ref{eval1})
shows that the eigenvector shape changes significantly, although
they remain spatially localized. Now the phases of
all the eigenvalues demonstrate significant dependence on $\theta$.
The crossings of the  $\arg \Lambda_n(\theta)$ curves demonstrate
that the localized eigenmode ``interacts'' with the modes of the linear spectrum.
First of all we note the
collision of the eigenvalues at $\arg \Lambda =\pi$ that
corresponds to the period-doubling bifurcation.
These bifurcations
are clearly seen if one compares the lines for $E_1=0.265$ and $E_1=0.269$
in Fig.~\ref{eval1}(a-b), however, they do not cause any instabilities.
After reaching some critical value of $E_1$, the localized eigenvalues
which previously
have resided in the complex plane collide with each other on the
real axis at ${\mbox {Re}} \Lambda>0$. The
further increase of $E_1$ leads to the fast exit of one of them out of the unit circle
and to the subsequent tangential bifurcation.
 Beyond this bifurcation point, the non-transporting limit cycle disappears.
For $E_1=0.27$ there exist four tangential bifurcations, two direct and two inverse, that
happen not far from the values $\theta=0$ and $\theta=\pm \pi$.
The respective unstable eigenvectors are spatially localized, as shown in
Fig.~\ref{eval1}(e).
The respective repeller also has been computed (not shown in the figure for the sake
of clearness), and its
eigenvalues join the attractor eigenvalues at the bifurcation point.
The transporting intervals $(\theta_1^-,\theta_2^-)$
and $(\theta_1^+,\theta_2^+)$ in Fig.~\ref{Vth} for $\kappa=0.09$ obviously
correspond to the windows in Fig.~\ref{eval1}(a-b) where the standing
fluxon does not exist. Similarly
to the dependencies shown in Fig.~\ref{fig3}, the bifurcation
points are preceded by the sharp singular-like growth
of the modulus of the unstable eigenvalue.

\begin{figure}[htb]
\includegraphics[width=1.0\columnwidth]{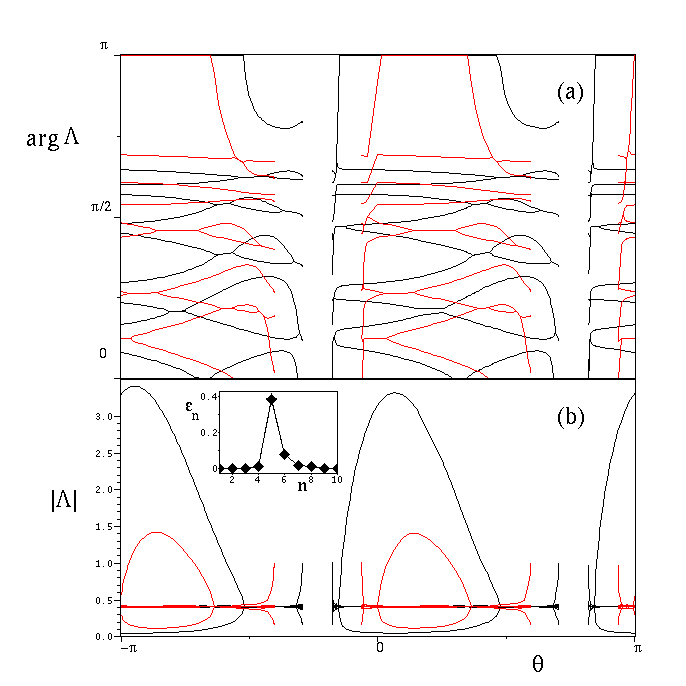}
\caption{(Color online). Phases (a) and moduli (b) of eigenvalues of
Floquet matrix
as a function of phase delay $\theta$ for $\kappa=0.12$, $E_1=E_2=0.25$ (red)
and $E_1=E_2=0.27$. The rest of parameters is as in Fig.~\ref{eval1}.
The inset shows the unstable eigenvector ${\mbox {Re}}~ \varepsilon$ ($+$) and
${\mbox {Im}}~ \varepsilon$ ($\blacklozenge$)
at $\theta=-3.03$ just after the period-doubling
bifurcation (see text for details).
}
\label{g6a}
\end{figure}

A different scenario is presented in Fig.~\ref{g6a}. As it has been
discussed in the previous subsection, the existence subspaces
of the transporting and non-transporting trajectories in the
parameter space $(\kappa,E_1,\theta)$ can take a peculiar and
complicated shape. This is also true for the discrete soliton ratchets with
 the spatial
 symmetry breaking studied in Ref.~\cite{cs-rs10pre} (see Fig.~12 therein).
If one stays above the critical dependence
$E_1^{(c)}(\kappa)$, it is possible to observe the existence of a
non-transporting limit cycle after the first tangential bifurcation as
shown in Fig.~\ref{fig4}. It is logical to investigate such
a case for different $\theta$'s when $E_1>E_1^{(c)}(\kappa)$. This exactly has been
done in Fig.~\ref{g6a} where the eigenvalue behavior has been
plotted as a function of $\theta$ for $E_1=0.25$ and $E_1=0.27$. On these
figures again four
tangential bifurcations are seen (two direct and two inverse), clearly pointing
out the windows where the directed fluxon motion takes place. These bifurcations
are characterized by sharp singular growth of the respective
$|\Lambda_n(\theta)|$ dependencies.
%

Apart from these bifurcations which are typical for the case
considered before (including Fig.~\ref{eval1}), the two large ``bubbles'' on the
dependence $|\Lambda_n(\theta)|$ can be noticed. They correspond
to the eigenvalue collisions on the real axis at $\mbox{Re}~\Lambda_n < 0$
and to the subsequent departure of one of them out of the unit circle. Thus, we observe
period-doubling bifurcations that make the mode-locked state ustable.
Along the interval $\pi \le \theta < \pi$ the four
period-doubling bifurcations take place: two direct and two inverse.
The destabilizing eigenvector is spatially localized as in the case
of tangential bifurcations (see the inset of Fig.~\ref{g6a}). Note that
the ``bubbles'' of the instable eigenvalues
increase when $E_1$ increases. The stable period-2
 (with $l=2$) limit cycles that appear after the bifurcation,
have been computed with the Newton method until
the next period-doubling bifurcation takes place. The emerging period-4
cycle
has been computed as well. In the next section, it will be shown that
in fact
 the cascade of period-doubling bifurcations takes place leading
to chaotic dynamics.

Now we can explain why the dependence $V(\theta)$ computed
in Fig.~\ref{Vth} for $\kappa=0.12$ has not two but four transporting
intervals. The old transporting intervals $(\theta_1^+,\theta_2^+)$
and $(\theta_1^-,\theta_2^-)$
now have non-transporting windows embedded within them. The borders
of these intervals
are associated with the depinning of fluxons and they, of course, do
not exactly coincide with the points of
period-doubling bifurcations of Fig.~\ref{g6a}. The actual depinning
takes place
after the sequence of the period-doubling bifurcations and transition
to chaotic dynamics.

\subsection{Dynamical properties of  depinned trajectories}

Thus, we have seen that the loss of stability of a pinned mode-locked
fluxon
takes place either due to a tangential bifurcation or after a
sequence of period-doubling
bifurcations. In this subsection, we focus on the fluxon dynamics
at the depinning threshold.

\begin{figure}[htb]
\includegraphics[width=1.0\columnwidth]{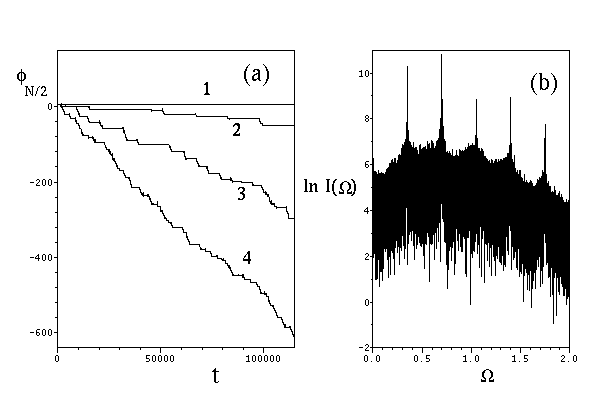}
\caption{Panel (a). Time evolution of  central junction phase
$\phi_{N/2}$ at $\kappa=0.12$, $E_1=E_2=0.27$, $\omega=0.35$, $\alpha=0.1$
for $\theta=-0.92462$ (1), $\theta=-0.9246$ (2), $\theta=-0.9245$ (3),
and $\theta=-0.924$ (4).
Panel (b) shows power spectrum for $\theta=-0.924$.
}
\label{fig7}
\end{figure}

Consider first the case of mode-locked state stability loss via
the tangential bifurcation as the phase shift $\theta$
varies. In this case, we consider the
already unstable mode-locked state with one of the eigenvalues
slightly out of the unit circle. After perturbing this state in the
direction of the unstable eigenvector, the directed fluxon propagation
begins. It has clear chaotic nature as shown in Fig.~\ref{fig7}(a).
%
The fluxon dynamics takes place according to the intermittency type-I
scenario: the junction phase (in this case the central one with $n=N/2$)
oscillates regularly around the equilibrium position as
shown by the plateaux of constant phase $\phi_{N/2}$
and then suddenly jumps in a certain direction of propagation
(sometimes after several chaotic back and forward jumps). The average
length of the regular (laminar) plateaux is inversely proportional
to the voltage drop. This length decreases as $\theta$ moves away
from the bifurcation point. Note the extremely sharp dependence of the modulus of the unstable
eigenvalue $|\Lambda_{unst}|$ as a function of $E_1$ and $\theta$ in
Figs.~\ref{fig3} and \ref{eval1}, respectively. Since
 the length of the laminar regime is proportional to $1/(|\Lambda_{unst}|-1)$,
it is necessary to change the respective parameter in the neighborhood of the
depinning transition with very small increments. Due to limited
computing power this procedure was not done and, as a result, the dependencies
in Fig.~\ref{Vth} seem to start ``out of nowhere''.

The  power spectrum of the junction's phase velocity  defined as
\begin{equation}\label{spectrum}
I(\Omega)=
\left | \int_{-\infty}^{+\infty} {\dot \phi}_{N/2}(t)e^{-i\Omega t}dt \right |^2,
\end{equation}
has been computed in Fig.~\ref{fig7}(a). The sharp peaks are located at the multiples
of the driving frequency: $\Omega=n\omega$, $n=1,2,\ldots$. These peaks
clearly illustrate the remnants of the mode-locked dynamics before the
tangential bifurcation.
The type-I intermittency in the depinning
transition for the single-harmonically driven kinks has been observed in
Ref.~\cite{mfmfs97prb}.

\begin{figure}[htb]
\begin{center}
\includegraphics[width=1.0\columnwidth]{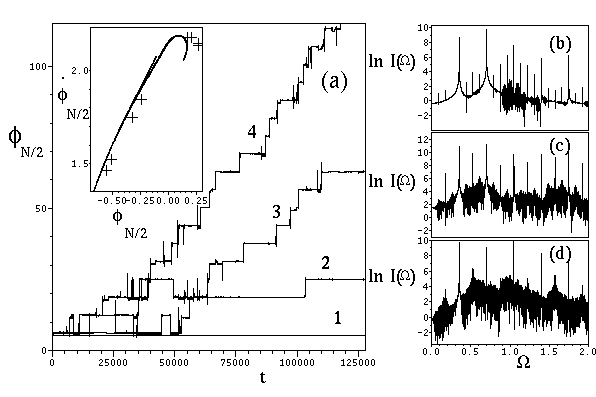}
\end{center}
\caption{Panel (a). Time evolution of the central junction phase
$\phi_{N/2}$ at $\kappa=0.12$, $E_1=E_2=0.27$, $\omega=0.35$
for $\theta=-2.08$ (1), $\theta=-2.11$ (2), $\theta=-2.115$ (3)
and $\theta=-2.12$ (4). The inset shows the Poincare
section for $\theta=-2.045$ ($+$) and $\theta=-2.08$ (dots).
Panels (b)-(d) show Fourier power spectra for $\theta=-2.05$ (b),
$\theta=-2.055$ (c) and $\theta=-2.08$ (d).
}
\label{fig8}
\end{figure}

Now we focus ourselves on the depinning scenario due to the period-doubling
bifurcations. Consider the
case depicted in Fig.~\ref{g6a} when the standing fluxon
loses its stability via the period-doubling bifurcation, for example,
at $\kappa=0.12$, $\theta \simeq -1.85$. The further decrease of $\theta$
leads to the sequence of period-doubling bifurcations where the
initially stable $2T$, $4T$, $\ldots$ cycles lose their stability.
A typical period-doubling route to chaos takes place and the
standing fluxon becomes quasiperiodic and later chaotic.
With decrease of $\theta$, the amplitude of chaotic oscillations
grows and the fluxon starts to propagate.
In Fig.~\ref{fig8}(a) the
typical trajectories after the depinning point are shown. Curve 1
corresponds to  non-transporting trajectories both regular and
chaotic. Since they overlap, the inset with the Poincare sections,
computed on the plane $\{ \phi_{N/2}, {\dot \phi}_{N/2} \}$ at every
time interval $t_n=nT$, $n=1,2,\ldots$ is presented.
Curves $2-4$ correspond to transporting
trajectories. The dynamical fluxon trajectories have the structure similar to the
case described in the previous paragraph. They consist of long
intervals, where the fluxon stays pinned, which
are interrupted by short chaotic bursts. During these
bursts (which occur at random) the fluxon jumps normally in the
direction specified by the asymmetry of the bias function $E(t)$.
%
Sometimes a series of forward and backward jumps takes place.
Again, the length of the pinned intervals decreases as $\theta$ moves
away from the critical depinning value. These dynamical
pictures bear a significant difference, although they look similar.
In the first case, the standing
fluxons are always regular and the transporting trajectory is chaotic
with long regular intervals where dynamics is very close to the periodic one
with the period $T$. In the depinning scenario, driven by the
period-doubling bifurcations, the chaotization happens before
depinning and the stationary intervals of the depinned trajectories
cannot be called laminar because their structure is in fact chaotic.

The power spectra in Fig.~\ref{fig8}(b-d) clearly indicate the
period-doubling route to chaos. Indeed, apart from the peaks at
$n\omega$, $n=1,2,\ldots$, one can clearly see the peaks
at $n\omega \pm m\omega/2$ and  $n\omega \pm m\omega/4$,
$m,n=1,2,\dots$, in Fig.~\ref{fig8}(b). As $\theta$ decreases
into the chaotic region, the peaks that correspond to the $\omega/4$
contribution get smeared out and only the $\omega/2$ peaks survive [see
panel (c)]. After the further decrease of $\theta$ the subharmonic peaks
become even less pronounced.

\section{Discussion and conclusions}
\label{conc}

In this paper, we have studied the fluxon dynamics in a highly discrete
annular Josephson junction array driven by the asymmetric periodic bias
current with the zero mean value. It is already well-known that this
bias (consisting of a cosine harmonic and its second overtone)
leads to a {\it directed} fluxon motion which is manifested by the non-zero
voltage drop. It is interesting to note that for the strongly
discrete JJA the ratchet transport is mainly chaotic, while the
non-transporting states are regular. Thus, the chaotic dynamics can
be identified experimentally directly from the IV curves.

We mainly focus on the depinning process of the fluxon in the
limit of the weak coupling between the neighboring junctions
($\kappa \ll 1$). It appears that the fluxon motion is possible in this
case provided the amplitude of the ac bias is sufficiently large.
The
existence diagram on the parameter plane $(\kappa,E_1)$ has been computed.
On this plane, the curve $E^{(c)}(\kappa)$ separates the area
where only mode-locked standing fluxons exist for any phase
shift $\theta$ from the area where moving fluxons can exist for
some values of $\theta$. In fact, this diagram can be treated as
the projection of a  more complicated
 three-dimensional parameter space $(\kappa,E_1,\theta)$ on the
 plane $(\kappa,E_1)$.

 We have investigated the depinning of the initially standing
mode-locked fluxon state not far from the critical
line $E_1^{(c)}(\kappa)$ by analyzing its Floquet spectrum.
The depinning occurs through chaotization of the mode-locked state.
The mechanisms of chaotization are diverse and depend on the
initial position of the non-transporting limit cycle
in the parameter space $(\kappa,E_1,\theta)$. The most common
depinning scenario occurs
through the type-I intermittency which evolves after the
tangential bifurcation destroys a standing fluxon limit cycle. This
scenario is ubiquitous when approaching the $E_1^{(c)}(\kappa)$
 curve from below by increasing the bias amplitude $E_1$.
Another scenario develops through a sequence of period-doubling
bifurcations. It happens slightly above the $E_1^{(c)}(\kappa)$ curve when
$\theta$  is varied.

Finally, we would like to point out some further research directions.
The main difficulty in studying the kink dynamics in strongly
discrete systems is the lack of an adequate analytical approximation.
As a result, we are forced to work with numerical methods.
The existing approximate theories are based on the perturbations of
the continuum models where kinks are treated as point particles
 (maybe with the internal mode taken into account) in the PN potential.
This approach works relatively well if $\kappa \gtrsim 1$ but
breaks if $\kappa \ll 1$. An important tool that could help to
obtain  the $E_1^{(c)}(\kappa)$ curve analytically is the Melnikov criterion.
However, the effectively use of it requires the explicit expressions for the
stable and unstable manifolds in order to build the Melnikov function.
For example, successful implementation
of the Melnikov method used in Ref. \cite{mc04prl,cm07prl} utilizes the
collective coordinate approximation which is not applicable in our case.
Performing the same for the case of small
$\kappa$ in our opinion is an important challenge.

In summary, we remark that the biharmonically driven ratchet effect in JJAs
is robust phenomenon that takes place even in the limit of very strong
coupling between the neighboring junctions.

\section{Acknowledgements}
\label{ack}
Financial support from DFFD (grant F35/544-2011) is acknowledged.


\end{document}